\documentclass{SciPost}

\hypersetup{
    colorlinks,
    linkcolor={red!50!black},
    citecolor={blue!50!black},
    urlcolor={blue!80!black}
}
\usepackage{amsmath}
\usepackage{physics}
\usepackage[bitstream-charter]{mathdesign}
\DeclareSymbolFont{usualmathcal}{OMS}{cmsy}{m}{n}
\DeclareSymbolFontAlphabet{\mathcal}{usualmathcal}

\fancypagestyle{SPstyle}{
\fancyhf{}
\lhead{\colorbox{scipostblue}{\bf \color{white} ~SciPost Physics }}
\rhead{{\bf \color{scipostdeepblue} ~Submission }}

\fancyfoot[C]{\textbf{\thepage}}
}

\begin{document}

\pagestyle{SPstyle}

\begin{center}{\Large \textbf{\color{scipostdeepblue}{
Phonon thermal Hall effect in quartz and its absence in silica}}}\end{center}

\begin{center}
\textbf{Yu Ling$^{1,2}$, Benoît Fauqué$^1$ and Kamran Behnia$^1$}
\end{center}

\begin{center}

{\bf 1} Laboratoire de Physique et d'Etude des Mat\'eriaux (CNRS) ESPCI Paris, PSL Universit\'e, 75005 Paris, France \\
{\bf 2} Wuhan National High Magnetic Field Center and School of Physics, Huazhong University of Science and Technology, Wuhan 430074, China
\\[\baselineskip]
$\star$ \href{mailto:kamran.behnia@espci.fr}{\small kamran.behnia@espci.fr}
\end{center}

\section*{\color{scipostdeepblue}{Abstract}}
\boldmath\textbf{The observation of a misalignment between the applied heat flux and the measured temperature gradient in insulating solids induced by magnetic field has become a subject of experimental investigation, theoretical speculation, and unsettled controversy. To identify the origin of this phonon thermal Hall ef
fect, we performed a comparative study of longitudinal and transverse heat transport in crystalline (quartz) and vitreous (silica) SiO$_2$ using identical experimental set-ups and thermometers. A finite signal was detected in the crystalline samples and none in the amorphous sample, within our resolution. The cleaner crystal exhibited a larger thermal Hall conductivity than the dirtier one, ruling out disorder as the driver of the effect. On the other hand, the amplitude of the transverse thermal resistivity is almost identical in the two crystalline samples (W$_{\perp}$/B$\approx 10^{-6}$ m.K.W$^{-1}$.T$^{-1}$). We show that in a phonon gas,  as in a molecular gas displaying the Senftleben-Beenakker effect, heat is conducted through two channels, and argue that a thermal Hall response is unavoidable whenever these channels differ both in entropy production and in their coupling to the magnetic field.  Under such conditions, the conserved energy current and the non-conserved entropy current cease to be parallel. Finally, the magnitude of the transverse thermal resistivity can be accounted for by a surprisingly simple picture. The heat flux induces a tiny drift velocity of the lattice nuclei, the magnetic field exerts a transverse Berry force on this drift, and this force is balanced by an entropic restoring force. }

\vspace{\baselineskip}

\vspace{10pt}
\noindent\rule{\textwidth}{1pt}
\tableofcontents
\noindent\rule{\textwidth}{1pt}
\vspace{10pt}


\section{Introduction}
\label{sec:intro}

Heat propagation differs dramatically between  crystalline  and amorphous solids \cite{Zeller1971,Vandersande01031986}. Periodicity allows the atomic vibrations of a crystal to become collective, allowing phonons to carry heat along many interatomic distances. In a glass, the absence of long-range order disrupts this mechanism. Not only the amplitude but also the temperature dependence of the thermal conductivity differ between glasses and crystals. Cooling a crystal below room temperature enhances thermal conductivity by increasing the phonon lifetime, limited by the presence of other phonons. In contrast, the thermal conductivity of glasses decreases with decreasing temperature following the reduction in the number of ephemeral thermally excited vibrations.

Thermal transport in crystals has been understood in great detail. In the intrinsic regime of phonon transport, starting above room temperature and down to a tenth of Debye temperature, a theoretical account of the experimental data has attained an impressive accuracy even in complex crystals \cite{Lindsay2013,mcgaughey2019}. In comparison, the conceptual modeling of thermal transport in glasses lags far behind. Nevertheless, following the pioneering experiments by Zeller and Pohl \cite{Zeller1971} and concepts put forward by Allen and co-workers \cite{allen1999diffusons} a phenomenological picture of the three different regimes of  thermal transport in glasses has gradually emerged \cite{DeAngelis03042019}. A recent unified approach \cite{Simoncelli2019}  has derived a single transport equation, based on the Wigner distribution, building a bridge between the Peierls limit \cite{Peierls1929} for anharmonic crystals and  the Allen–Feldman limit \cite{allen1999diffusons} for harmonic glasses. 

In this context, the recent experimental observation of a phonon thermal Hall effect \cite{Strohm2005} in numerous crystalline insulators \cite{Ideue2017,Sugii2017,Li2020,Grissonnanche2020,Boulanger2020,Akazawa2020,Sim2021,Chen2022,Uehara2022,Jiang2022,Li2023,Chen2024,Chen2024-2,Ataei2024,Meng2024,sharma2024phonon,Li2025,xiang2025arxiv}, including in simple elemental insulators \cite{Li2023,lishi2025} has been unexpected. The experimental observation consists of a small, yet detectable field-induced misalignment between the heat current density and the temperature gradient vectors \cite{Behnia2025}. Most theories \cite{Sheng2006,zhang2010topological,Qin2012,Chen2020,Flebus2022,Guo2022,Mangeolle2022} neglect interactions between phonons. Some of them highlight the role of disorder \cite{Chen2020,Flebus2022,Guo2022}. 

\begin{figure*}
\centering
    \includegraphics[width=\textwidth]{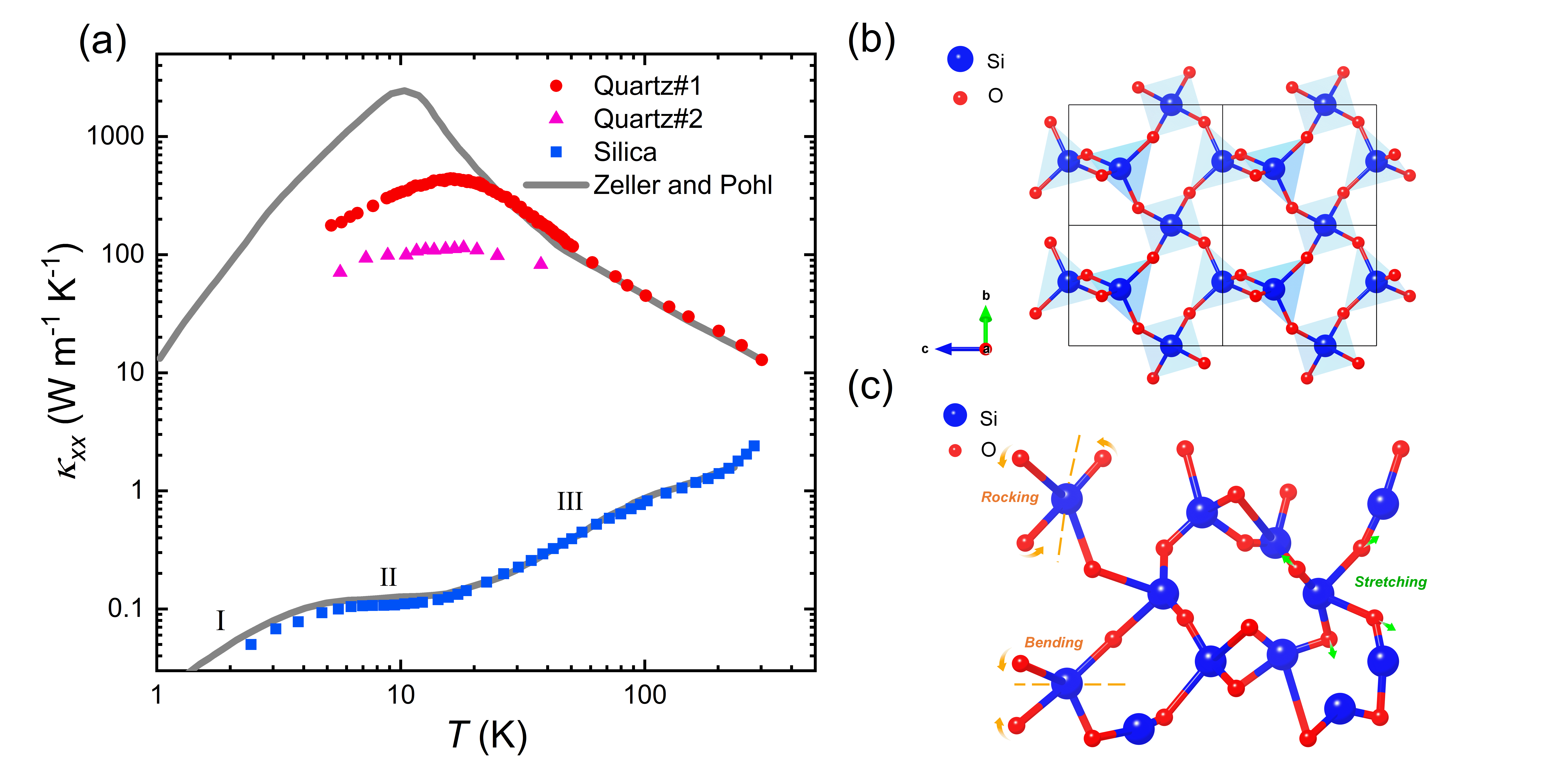}
    \caption{\textbf{ Comparison of longitudinal thermal conductivity and atomic structure of quartz and silica.} a) Measured $\kappa_{xx}$ for the two quartz samples is shown by red circles (Quartz \#1) and pink triangles (Quartz \#2). Blue squares represent the measured thermal conductivity of the silica sample. Gray solid lines represent the $\kappa_{xx}$ values reported in \cite{Zeller1971} for quartz and silica. b) Side view (from a-axis) of the periodic lattice of quartz. Silicon atoms and oxygen atoms are shown as blue and red spheres. SiO$_4$ tetrahedra are highlighted in light blue. c) Schematic diagram of the disordered  network of corner-sharing, distorted SiO$_4$ tetrahedra in silica. Curved orange arrows indicate possible motions of oxygen atoms and green arrows indicate stretching motions between oxygen and silicon atoms \cite{taraskin1997silica}.}
    \label{fig:1}
\end{figure*}

In this paper, we report on a study of longitudinal, $\kappa_{xx}$, and transverse, $\kappa_{xy}$, thermal conductivity in quartz and silica. These two solids share the same structural building blocks, namely the SiO$_4$ tetrahedra, but differ in their arrangement (see Fig. \ref{fig:1}). The experiments were carried out under identical conditions on  two quartz crystals and one silica glass.  In the two crystals, we detect a finite transverse temperature gradient induced by the magnetic field. The cleaner crystal, with a higher longitudinal thermal conductivity, has also a larger thermal Hall conductivity. In contrast,  the glassy sample does not display a detectable Hall signal. An  experimental upper bound can be placed on the amplitude of any transverse response in silica. Our study confirms that disorder weakens (instead of amplifying) the thermal Hall response and  identifies crystallinity as a necessary ingredient for the emergence of a thermal Hall response. 

We will argue that our results are compatible with the idea of phonon-phonon collisions  \cite{Behnia2025} as the driver of the thermal Hall effect by comparing the phonon gas with a real gas. The so-called Senftleben-Beenakker (SB) effect \cite{Beenakker1962,kagan1967kinetic} in molecular gases arises due to the  field-induced modification of the molecular collision rate \cite{kagan1967kinetic}. It turns the thermal conductivity of gases into a tensor with off-diagonal components,  odd in magnetic field. We put forward an analogy between a phonon gas and a molecular gas. In both cases, the magnetic field influences entropy-producing events, namely collisions between particles and this makes a field-induced misalignment between the entropy flux and the thermal energy flux  unavoidable.

Finally, we  find that the amplitude of the transverse thermal resistivity is of the order of magnitude expected if the tiny drift velocity of atomic nuclei induced by thermal energy flux generates a field-induced Berry force on the nuclei countered by an entropic force.

\section{Results}
\subsection{Longitudinal thermal transport}
Fig. \ref{fig:1}a shows the temperature-dependence of longitudinal conductivity, $\kappa_{xx}$, in our samples compared to what was reported by Zeller and Pohl \cite{Zeller1971}. Our silica data superpose with theirs. The peak thermal conductivity is lower in our quartz samples, which are dirtier and thinner. However, above 300 K, in the intrinsic regime, our crystal data points match theirs too.

A rough account of the temperature dependence of the thermal conductivity is given by the kinetic formula $\kappa=\frac{1}{3}C_v v_s l_{ph}$, where $C_v$ is the heat capacity per volume, $v_s$  the average sound velocity and $l_{ph}$ the phonon mean free path. Upon cooling from room temperature, thermal conductivity in quartz  increases due to the reduction in the frequency of Umklapp scattering events. At sufficiently low temperature,  $l_{ph}$  approaches the sample size and saturates. In this regime, boundary scattering dominates and the temperature dependence of specific heat sets the temperature dependence of $\kappa_{xx}$. The competing thermal variations of the phonon density and mean free path lead to a peak in  thermal conductivity. The position and the amplitude of the $\kappa_{xx}$ peak imply that Quartz \#2 is dirtier than Quartz \#1. 

It is worth mentioning that we found a room temperature thermal conductivity of 13 W.K$^{-1}$.m$^{-1}$ in quartz.  This value is almost identical to what was measured by Zeller and Pohl \cite{Zeller1971} and by Kanamori and co-workers \cite{Kanamori1968}. But it is slightly larger than the value calculated by Mizokami \textit{et al.}  (10.8$\pm$0.1 W.K$^{-1}$.m$^{-1}$) using first-principles calculations, as well as solving Boltzmann transport equation \cite{Mizokami2018}. 

\begin{figure*}[ht!]
\centering
    \includegraphics[width=\textwidth]{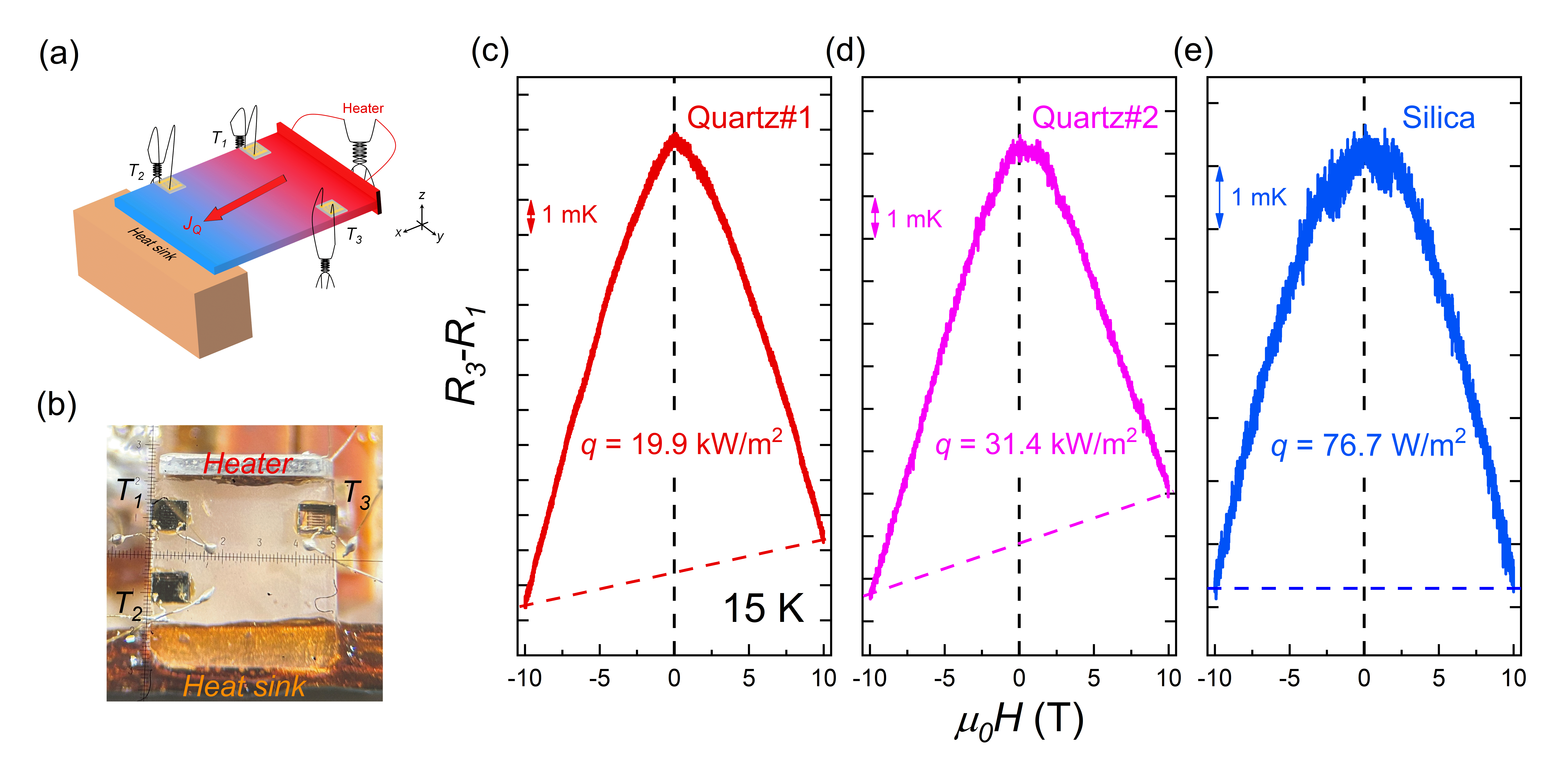}
    \caption{\textbf{Comparison of raw data in quartz and in silica.} a) Sketch of the experimental set-up with the sample sandwiched between a heater and a cold finger supporting three Cernox thermometers. b) A photograph of a sample with thermometers. c-e) Raw data for Quartz \#1 (c), Quartz \#2 (d) and silica (e) at 15 K. $R_1$ and $R_3$ are the resistance of two Cernox thermometers labeled 1 and 3 and monitoring the transverse temperature difference. The dashed lines highlight the presence of asymmetry due to an odd response in quartz and its absence in silica.}
    \label{fig:2}
\end{figure*}
As seen in Fig. \ref{fig:1}a, the magnitude and the temperature dependence of $\kappa_{xx}$ in silica are very different. It decreases monotonically with cooling and there is no regime dominated by phonon-phonon scattering in which thermal conductivity increases with cooling. One usually distinguishes between three regimes of heat transport. At low temperature (regime I) the main carriers of heat are `propagons', low frequency, acoustic-like vibrational modes, associated with bending or rocking motion of Si-O bonds (Fig. \ref{fig:1}c) \cite{allen1999diffusons,taraskin1997silica}.  Compared to the acoustic phonons of quartz, they travel  over shorter distances.  In regime II, $\kappa_{xx}$ exhibits a plateau, concomitant with a hump in $C/T^3$ often referred to as the boson peak. The disordered atomic structure, associated with spatial fluctuations in density and elastic constants, can produce Rayleigh scattering of propagons. Above the plateau (regime III), the contribution of `diffusons', vibrations with a mean free path comparable to their wavelength, and unable to propagate, begin to become significant.
\subsection{Transverse thermal transport}
Having briefly contrasted longitudinal heat transport in quartz (with heat carriers  and their scatterers clearly identified), and silica (where the identity of both is fuzzier), let us now put under experimental scrutiny their transverse thermal conductivity.

We used a one-heater-three thermometers setup (Fig. \ref{fig:2}a,b) to measure thermal Hall conductivity $\kappa_{xy}$. In both quartz crystals,  we applied the heat current along the $[11\bar{2}0]$ direction and the magnetic field along $[0001]$ direction. The longitudinal temperature difference  was measured along $[11\bar{2}0]$ and the transverse along $[1\bar{1}00]$. Fig. \ref{fig:2}c-e compares the raw data for Quartz\#1, Quartz\#2 and Silica samples at 15 K. What is shown is the field-induced difference in the resistance of two Cernox thermometers  monitoring the transverse temperature difference. The three panels show the results obtained when the magnetic field was swept from -10 T to 10 T. The signal contains an antisymmetric component (highlighted by dashed lines) in Quartz\#1 and  Quartz\#2 but not in Silica. The symmetric component is mostly due to the magneto-resistance of the Cernox thermometers (which are not identical). It is also partly due to a field-induced change in $\kappa_{xx}$ combined with an unavoidable misalignment between lateral positions of the two sensors. The antisymmetric component, odd in magnetic field is the genuine signature of the thermal Hall effect.

\begin{figure*}
\centering
    \includegraphics[width=\textwidth]{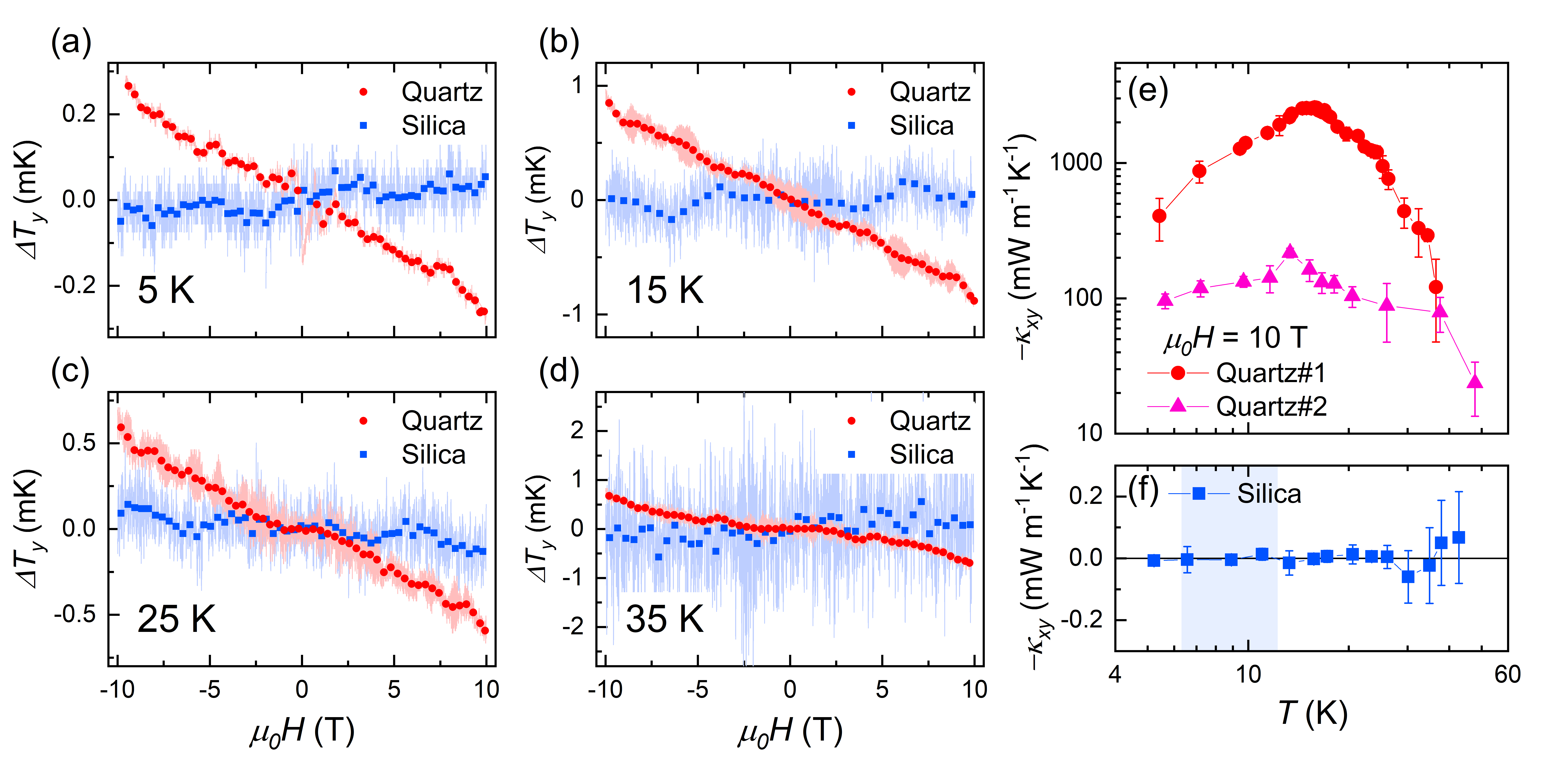}
    \caption{\textbf{The transverse temperature difference and the thermal Hall conductivity.} a-d) $\Delta T_y$ extracted from the raw data. Light red and light blue lines represent $\Delta T_y$ in quartz and silica. Red circle and blue square symbols represent the average of 200 measurements. An odd response is detectable in quartz and absent in silica. e) Temperature-dependence of $-\kappa_{xy}$ in Quartz \#1 (red circles) and in Quartz \#2 (pink triangles), plotted on a logarithmic scale. f) Temperature-dependence of $-\kappa_{xy}$ of silica, plotted in a linear scale. The  plateau region of $\kappa_{xx}$ is highlighted in light blue.}
    \label{fig:3}
\end{figure*}

The field dependence of the transverse temperature gradient, $\Delta T_y$, at different temperatures is shown in Fig. \ref{fig:3}a-d. Note that the symmetric component has been subtracted. In quartz, a finite $\Delta T_y$, odd and linear in the magnetic field, can be clearly detected, despite its small amplitude (less than 1 mK). On the other hand, within our signal-to-noise resolution, there is no detectable signal in silica at any temperature.  Fig. \ref{fig:3}e shows the temperature dependence of $\kappa_{xy}$  at 10 T in the two quartz samples. Note that the cleaner sample, the one with a larger $\kappa_{xx}$, has  a much larger $\kappa_{xy}$ too.  Fig. \ref{fig:3}f shows data points for silica together with an estimated margin of error.

\begin{figure*}
\centering
    \includegraphics[width=\textwidth]{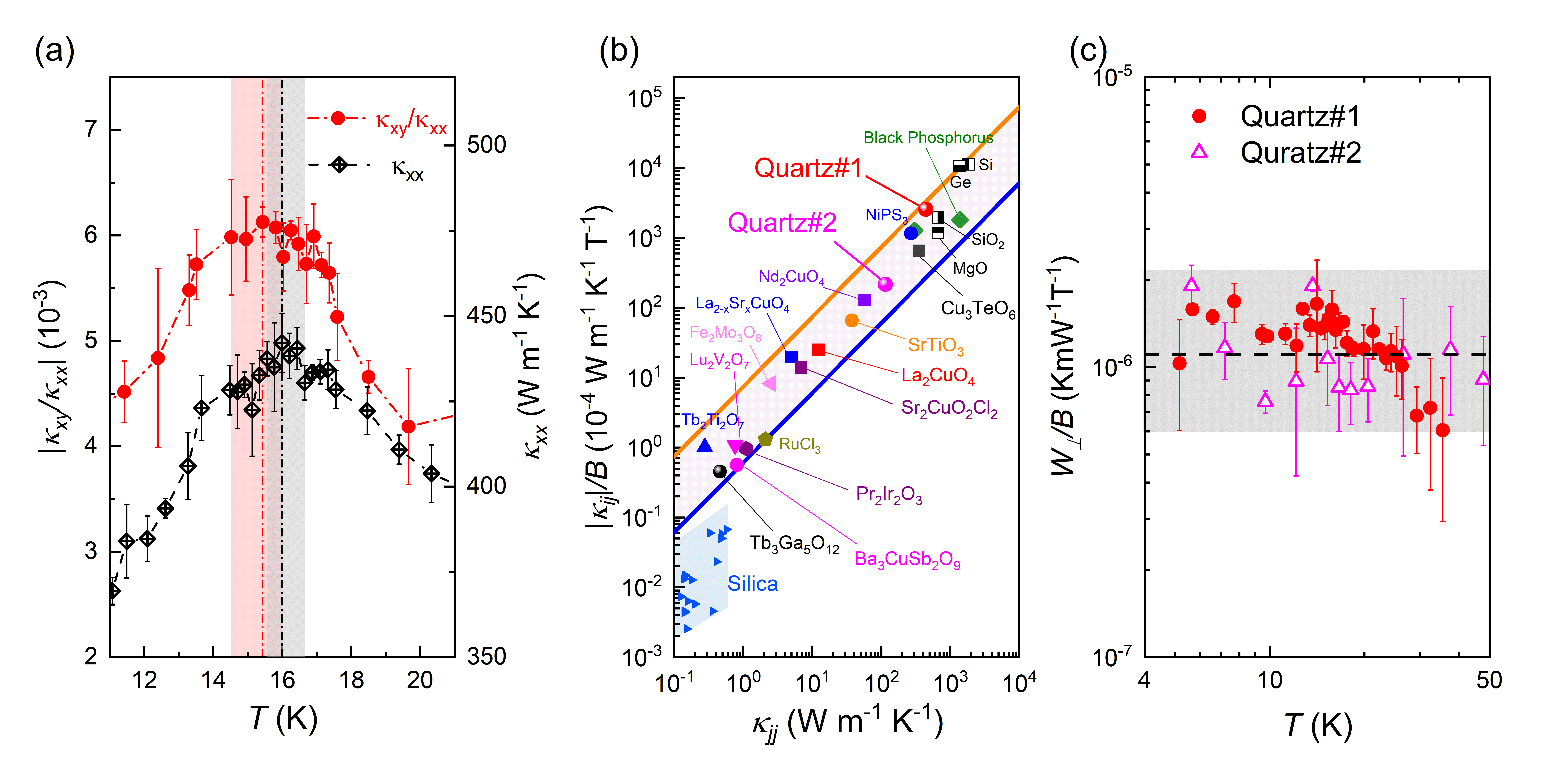}
    \caption{\textbf{Thermal Hall angle and thermal Hall resistivity.} a) The thermal Hall angle, |$\kappa_{xy}/\kappa_{xx}$| (red circles), and the longitudinal thermal conductivity, $\kappa_{xx}$ (black diamonds), near their maximum values. They peak at 15.4 K and 16 K, respectively. Light red and gray vertical stripes highlight the vicinity and the partial overlap of the two peaks. b) Maximum $|\kappa_{ij}|/B$ as a function of maximum $\kappa_{jj}$ in different insulators. Our  data points for Quartz \#1 and Quartz \#2 samples are shown by red and pink circles. Data points for silica at different temperatures are also shown as blue triangles in the light-blue region. They remain below our margin of error. c) Thermal Hall resistivity as a function of temperature in the two quartz samples. Within our experimental uncertainty, there is no detectable temperature or sample dependence. }
    \label{fig:4}
\end{figure*}
Fig. \ref{fig:4}a compares the temperature dependence of $\kappa_{xx}$ and the thermal Hall angle, $\frac{\kappa_{xy}}{\kappa_{xx}}$, in the cleaner quartz sample.  The first peaks at 16 K. The second attains a maximum of $6 \times 10^{-3}$ at 15.4 K. This difference in peak temperatures is within our margin of error. In this respect, quartz is similar to many other insulators in which a thermal Hall signal has been observed.  This universality, first noticed in  \cite{Li2020,Li2023}, was argued \cite{Behnia2025} to indicate a link between Normal phonon-phonon collisions and the physics underlying the thermal Hall effect.  Another universal feature is shown in Fig. \ref{fig:4}b. It represents the maximum $\frac{1}{B}\frac{\kappa_{xy}}{\kappa_{xx}}$ in different solids as a function of their peak $\kappa_{xx}$. The plot suggests a universal bound to the maximum thermal Hall angle. One can see that our data for the two quartz crystals is compatible with this bound.

Fig. \ref{fig:4}c, shows the thermal Hall resistivity of the two samples, the transverse temperature gradient divided by the longitudinal heat density flux:
$W_{\perp}=\frac{\nabla_yT}{q_x}$. As seen in the figure, the data for the two samples nearly overlap. In our temperature range of investigation 5 $K$ < T < 40 $K$, we cannot resolve any detectable temperature variation.  The scattering of data points yields $6 \times 10^{-7}$ m.K.W$^{-1}.$ T$^{-1}< $W$_{\perp}$/B$< 2 \times 10^{-6}$ m.K.W$^{-1}.$ T$^{-1}$. 

Below, we will come back to the significance of this observation.

\section{Discussion}
\subsection{Disorder and transverse thermal conductivity}
Our results on the two quartz samples indicate that the amplitude of the thermal Hall effect anti-correlates with disorder. 

A similar conclusion has been reached in several other cases. In strontium titanate, the scrutiny of the correlation between the amplitudes of transverse and longitudinal thermal conductivities  \cite{xiang2025phononthermalhalleffect} shows that samples with a lower $\kappa_{xx}$  display a smaller or undetectable $\kappa_{xy}$. It has also been noticed \cite{Jiang2025} that the absence of a measurable $\kappa_{xy}$ in La$_2$CuO$_4$ was  reported on a sample \cite{Hu2025} significantly less conductive than the one \cite{Grissonnanche2020} in which a finite $\kappa_{xy}$ was detected. Another interesting case is NiPS$_3$ \cite{Meng2024}, a honeycomb zigzag antiferromagnet. The amplitude of $\kappa_{xy}$ is lower in samples hosting structural domains reducing the intrinsic magnetization anisotropy.  

In contrast to these cases, the study of thermal Hall effect in La-doped Sr$_2$IrO$_4$ \cite{Ataei2024} concluded that  the thermal Hall effect of the material is caused by the scattering of phonons by impurities. Note, however, that in contrast to  our study and the cases listed above, the comparison was not made between  samples with identical stoichiometry and different impurity concentration, but between compounds with distinct chemical composition, different ground state, and presumably different phonon spectra. 

As for theories, phonon-impurity scattering or the side-jump \cite{Guo2022} version of it, has been proposed by several authors. Chen \textit{et al.} \cite{Chen2020}  proposed a scenario in which flexoelectricity leads to strong skew scattering of phonons. Flebus and MacDonald \cite{Flebus2022} pointed out that in ionic solids, charge defects can cause skew scattering and a  thermal Hall effect. Given the heaviness of the nuclei compared to electrons, the magnitude of the experimentally observed signal was often considered too large for an intrinsic response in a gas of neutral quasi-particles.

What was forgotten was the fact that \textit{real} neutral molecular gases host a detectable and well-understood thermal Hall response and this understanding does not make any reference to impurities. Let us briefly recall this half-forgotten episode of twentieth century physics. 
 
\subsection{Transverse thermal conductivity in a molecular gas}
In 1930, Senftleben \cite{Senftleben1930} found that the thermal conductivity of gaseous oxygen changed by a few per mille in a magnetic field. For several decades, it was thought that this is an exclusive property of paramagnetic gases. In 1962, Beenakker \textit{et al.} \cite{Beenakker1962} observed a similar field dependence of thermal conductivity in polyatomic gases. 

The physics behind the Senftleben-Beenakker effect \cite{Beenakker1970} is both simple and subtle.  The thermal energy of a  molecule is partially rotational. The collision probability for non-spherical molecules is not isotropic in space. Collisions tend to align them with each other. Since both diamagnetic and paramagnetic molecules have residual magnetic moments (albeit with different signs and amplitude), a finite magnetic field will generate a precession of the molecular angular momentum of the molecule. A  temperature gradient and magnetic field compete with each other leading to a field dependence of the thermal conductivity.

In this context, the existence of a transverse thermal conductivity in gases  was independently predicted  by Kagan $\&$ Maksimov \cite{kagan1967kinetic}, by Knaap $\&$ Beenakker \cite{KNAAP1967643}, and by McCourt $\&$ Snider \cite{McCourt1967}. It was subsequently verified experimentally \cite{HERMANS196781,hermans1970transverse}. The origin of this chiral response resides in the skewness of intermolecular collisions induced by the magnetic field. During a collision the angular-momentum of the two-molecule system has opposite orientations for molecules coming from the left and the right and this leads to an odd response as a function of the magnetic field. A detailed account of this effect requires  the Waldmann—Snider kinetic formalism, an extension of the Boltzmann formalism \cite{Mccourt1990}.

Let us distinguish between two channels of transport. Each molecule can carry its rotational and translational energy along the hot-cold axis.  Molecules with a large angular momentum have a larger effective collisional cross section. In contrast, the collision probability does not enhance with increasing linear momentum. Therefore, the rotational component of the thermal energy flux is more  entropized than the translational component.  One can invoke a two-channel conduction with a fuzzy boundary.  The first channel of heat flux, $\mathbf J_{\mathrm{tr}}$, consists of fast-moving molecules whose translational kinetic energy dominates. There is a second channel, $\mathbf J_{\mathrm{rot}}$, of slow-moving molecules with high angular momentum. This second channel heavily experiences the torque  of the field. The precession induced by magnetic field influences the second channel (which has a higher rate of entropy to energy content) much more than the first. 

We will see below that the distinction between the two channels can be identified as a generic driver of the thermal Hall response. 

Let us now return to the phonon gas.

\subsection{Callaway picture and the two components of phonon flow}
In 1959, Callaway \cite{Callaway1959} proposed a model  to describe heat conduction in crystalline solids, in which a distinction was made between phonons based on how they interact with each other. Normal (N)  phonon-phonon collisions conserve crystal momentum. They redistribute energy among phonons without directly causing thermal resistance. In contrast, Umklapp (U)  collisions do not conserve crystal momentum. They `flip'  the wavevector across a Brillouin-zone boundary. Thermal resistance is mainly caused by these collisions. In Callaway's model, N-events are not directly resistive yet they indirectly affect thermal conductivity by redistributing the phonon distribution, which in turn influences U-events.

This leads to  two  phonon reservoirs loosely separated according to the nature of their dominant scattering channel. The first reservoir (let us call it $L$) contains  $q < G/2$ phonons. They lie sufficiently close to the zone center such that no three-phonon collision can produce a reciprocal lattice vector $\mathbf{G}$. Those with $q > G/2$ (reservoir $H$) can undergo Umklapp in a three-phonon event. N events conserve crystal momentum and produce no entropy.  U events do not conserve crystal momentum and are the primary source of entropy production.

In Callaway's picture, the Boltzmann equation for phonon distribution $n(\mathbf{q},\omega)$ becomes:

\begin{equation}
  \left.\frac{\partial n}{\partial t}\right|_{\mathrm{drift}}
  = -\frac{n - n_{\lambda}}{\tau_{N}}
    - \frac{n - n_{0}}{\tau_{R}},
\end{equation}

Here, $n_{0}(\omega)$ is the equilibrium Bose--Einstein distribution, $n_{\lambda}(\omega)$ is a \emph{shifted} (drifting) Bose--Einstein distribution. $\boldsymbol{\lambda}$ is a drift wavevector, $\tau_{N}$ is the N-process relaxation time, and $\tau_{R}$ is the resistive relaxation time (Umklapp, defect, boundary).

A phonon in Reservoir $L$ can undergo a \emph{chain} of N-events that progressively accumulate wavevector until the threshold $G/2$ is crossed:

\begin{equation}
  \mathbf{q}_{1} + \mathbf{q}_{2} \to \mathbf{q}_{3}, \qquad |\mathbf{q}_{3}| > G/2.
\end{equation}

Once promoted into Reservoir $H$, the phonon becomes vulnerable to Umklapp. For a phonon in reservoir $H$ to re-enter the drifting population, it requires an N-event decreasing its wavevector below $G/2$, such that $\mathbf{q}_{1} + \mathbf{q}_{2} \to \mathbf{q}_{3}$ with $ \qquad |\mathbf{q}_{3}| < G/2$. This process is restricted by the scattering phase space. In three dimensions, the density of phonon states grows as $q^{2}$, so the volume of phase space above $G/2$ exceeds that below it. A random N-scattering event therefore preferentially pushes phonons \emph{toward} high $q$, making the downward crossing geometrically less probable than the upward crossing. Therefore:
\begin{equation}
  \Gamma_{L \to H} \;>\; \Gamma_{H \to L},
\end{equation}

Thus, there is an asymmetric leak between the two reservoirs. Low-$q$ phonons feed entropy production more efficiently than high-$q$ phonons are recycled into the drifting population. In the presence of a finite heat flux, both reservoirs contribute to the energy flux, but their entropy production is different.

The coupling of the two reservoirs to magnetic field is not identical. This is true irrespective of microscopic details, provided that the coupling has any $\omega$ or $q$ dependence. A specific proposal is  the acquisition of an unavoidable geometric phase by the Born-Oppenheimer approximated atomic wavefunctions in  a finite magnetic field  \cite{Scmelcher1988,Yin1994,Resta_2000,Culpitt2021}. This yields a geometric [Berry] phase to phonons \cite{Saito2019,Behnia2025}, weighing on pseudo-momentum balance rules \cite{Behnia2025}.

We will see below that, even without any assumption about the microscopic origin, an unequal coupling between the magnetic field and the two components of the heat flux is sufficient for the emergence of a transverse response.

\begin{figure*}
\centering
    \includegraphics[width=0.75\textwidth]{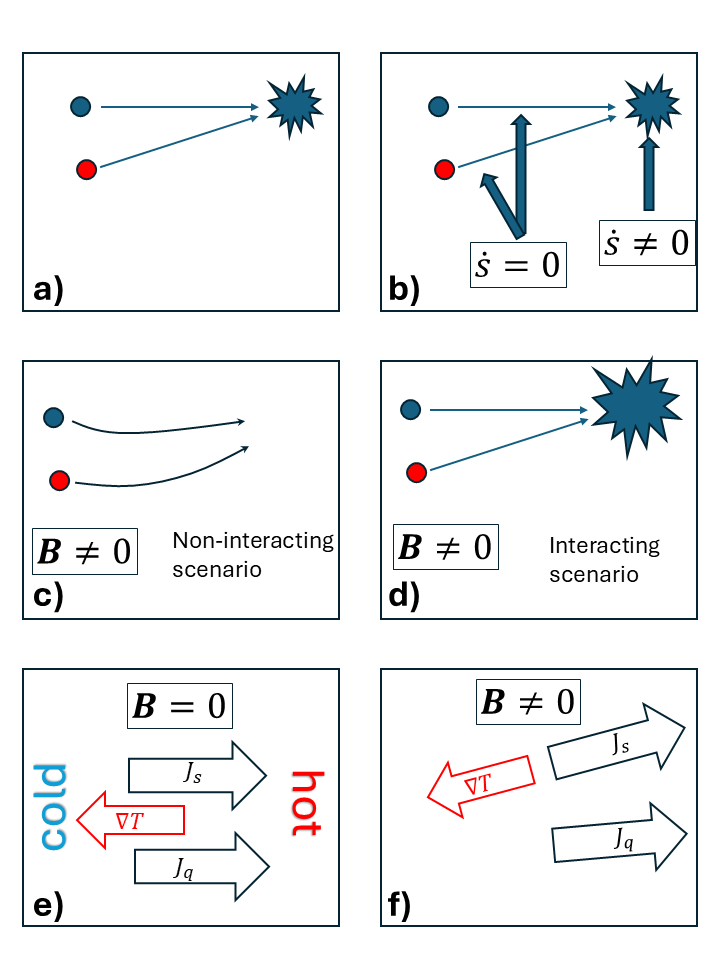}
    \caption{\textbf{Field-induced twist angle between thermal energy flux and thermal gradient via the coupling between magnetic field and the entropy flux.} a) Particles of gas travel along a trajectory (shown by arrows) until a collision event (shown by an explosion icon). b) Entropy is produced during collisions but not along the inter-collision trajectory. c) Non-interacting scenarios invoke bending of trajectories by the magnetic field. d) Interacting scenarios invoke the influence of magnetic field on collision events, which are associated with production of entropy. e) At zero magnetic field,  a temperature gradient vector generates opposite flows of thermal energy $J_q$ and entropy $J_s$ oriented along itself and parallel to each other. f) At finite magnetic field, in an interacting scenario, the magnetic field by coupling to $J_s$ twists it. The temperature gradient vector remains anti-parallel to $J_s$ and is no more parallel to $J_q$. }
    \label{Fig-vectors}
\end{figure*}

\subsection{Misalignment between thermal energy flux and entropy flux}
Consider now a medium in which the total heat  current is carried by two distinct channels:

\begin{equation}
\mathbf J_{\mathrm{q}} = \mathbf J_{\mathrm{1}} + \mathbf J_{\mathrm{2}},
\end{equation}

In a molecular gas, $\mathbf J_1 = \mathbf J_{\mathrm{tr}}$ and $\mathbf J_2 = \mathbf J_{\mathrm{rot}}$; in a phonon gas,  $\mathbf J_1 = \mathbf J_{L}$ and $  \mathbf J_2 =\mathbf J_{H}$. Let us now demonstrate that:
\begin{quote}
\emph{If the two energy-carrying components do not produce entropy equally, and if a magnetic field does not affect them identically, then the total energy flux and the total entropy flux cease to be parallel.}
\end{quote}

Irrespective of microscopic details, the combination of the two conditions leads to a thermal Hall effect, a misalignment between energy flux and entropy flux \cite{Behnia2025}. We start with three observations, in the framework of extended irreversible thermodynamics \cite{Jou2009}:

\begin{enumerate}
    \item Entropy is not conserved, but energy is conserved.
    \item The two components of energy flux are not identical in entropy production along their propagation.
    \item A magnetic field does not affect the two components identically.
\end{enumerate}

The first observation can be stated in the following way. Let \(\varepsilon\) be the energy density and $\mathbf J_{\mathrm{q}}$ the energy flux. Then energy conservation implies:

\begin{equation}
  \frac{\partial \varepsilon}{\partial t} + \nabla \cdot \mathbf J_{\mathrm{q}} = 0.  \end{equation}

Assigning \(s\)  to the entropy density, \(\sigma\) to the entropy production density and $\mathbf J_{\mathrm{S}}=\mathbf J_{\mathrm{q}}/T$  to the entropy flux, the non-conservation of entropy leads us to: 

\begin{equation}
\frac{\partial s}{\partial t} + \nabla \cdot \mathbf J_{\mathrm{S}} = \sigma.
 \end{equation}

Now, let us suppose that the magnetic field, an axial vector, generates symmetry-allowed transverse corrections to each of the two fluxes, which are polar vectors. The transverse component in the energy flux can still be decomposed into two components:

\begin{equation}
\mathbf J_{\mathrm{q},\perp} = \mathbf{J}_{q1,\perp} + \mathbf{J}_{q2,\perp}.
 \end{equation}

While both channels contribute to \(\mathbf J_{\mathrm{q}}\), they contribute with different weights to \(\mathbf J_{\mathrm{S}}\). Moreover,  the magnetic field affects the two channels differently. The weighted sum defining entropy transport cannot remain parallel to the unweighted sum defining energy transport. Therefore:

\begin{equation}
\mathbf{J}_{s,\perp} = \frac{\mathbf{J}_{q1,\perp}}{T_1} + \frac{\mathbf{J}_{q2,\perp}}{T_2} + \Delta \mathbf{J}_{coll.}(B)
\end{equation}

Here, the term $\Delta \mathbf{J}_{coll.}(B)$ refers to the entropy flux carried  by the internal degrees of freedom of the particles capable of coupling with the magnetic field. The two preceding equations imply: 

\begin{equation}
    \mathbf J_{\mathrm{S}} \not\parallel \mathbf J_{\mathrm{q}}.
\end{equation}

Thus, in the presence of two distinct components of heat flow, which do not couple identically to the magnetic field and do not produce entropy identically, a misalignment between the two fluxes is unavoidable (See Figure \ref{Fig-vectors}). 

\subsection{Field-induced transverse thermal resistivity}
A heat flux density of $J_{\mathrm{q}}^x$ generates a drift velocity of the elastic energy density. This can be written as: 

\begin{equation}
J_{\mathrm{q}}^x=\rho v_s^2 v_d
\label{heat flux}
\end{equation}
Here, $\rho$ is the mass density and $v_s$ the speed of sound. $\rho v_s^2$ is of the order of the second-order elastic constants of the solid. $v_d$ is  the drift velocity of the elastic medium. 

Equation \ref{heat flux} is based on the assumption that the excess phonon flow from hot side to cold side ends up generating a tiny asymmetry in the average velocity of each atom inside its potential well. Now, each atom consists of a nucleus and its surrounding electronic cloud. Let us suppose that magnetic field and this drift velocity lead to a Berry force as in the case of molecules \cite{Peters2022,Peters2023}: 

\begin{equation}
F_{\perp}\approx\frac{J_{\mathrm{q}}^x}{\rho v_s^2} Ze|B| 
\label{Berry}
\end{equation}

Here, $Z$ is the atomic number. In adiabatic conditions, this force will be countered by [a transverse flow of phonons producing] a thermal force equal to $k_B \nabla_yT$. Note that such a simple expression for an entropic force assumes that the nuclei are in local thermodynamic equilibrium. This appears reasonable given that the applied heat flux density $J_{\mathrm{q}}^x$ < 100 W.m$^{-2}$ corresponds to a drift velocity of $v_d$ < $10^{-6}$ m.s$^{-1}$. This is almost ten orders of magnitude smaller than the sound velocity or the atomic thermal velocity of atoms. There is enough time for a thermodynamic compensation of the Berry force.

Assuming $F_{\perp}\simeq k_B \nabla_yT$, equation \ref{Berry} leads to the following expression for transverse thermal resistivity:

\begin{equation}
    W_{\perp}\equiv\frac{\nabla_yT}{J_{\mathrm{q}}^x}\approx\frac{Ze|B|}{k_B\rho v_s^2}
    \label{TRH}
\end{equation}

Note that this transverse thermal resistivity is related to the thermal Hall conductivity through matrix inversion:

\begin{equation}
    W_{\perp}\equiv\frac{\kappa_{xy}}{\kappa_{xx}^2+\kappa_{xy}^2}
\end{equation}
Equation \ref{TRH} will become more transparent, by replacing the mass density by 
$\rho= Am_pa^{3}$, where $A$ is the atomic mass number, $m_p$ the proton mass and $a$ the average interatomic distance. With $\frac{A}{Z}\approx 2 $, one finds:

\begin{equation}
   \frac{W_{\perp}}{B} \approx  \frac{e}{2k_Bm_p}\frac{a^3}{v_s^2}
    \label{TRH2}
\end{equation}

The sound velocity in SiO$_2$ is 5.8 km/s. The interatomic distance is 2.32 \AA. Substituting these two material-dependent parameters in equation \ref{TRH2}, we find  W$_{\perp}$/B=1.3$\times 10^{-6}$m.K.W$^{-1}$.T$^{-1}$. 

As seen in Fig. \ref{fig:4}c, the experimentally measured values of $W_{\perp}/B$ are close to this. 
Astonishingly, despite the presence of several questionable approximations in the chain of reasoning, the final simple expression, Equation \ref{TRH2} matches the order of magnitude of the measured signal. Moreover, it is compatible with the absence of any detectable temperature dependence of W$_{\perp}$. Future studies will tell to what extent this agreement is an accident.

Note that a temperature-independent W$_{\perp}$ is compatible with the $\kappa_{xy}\propto\kappa_{xx}^2$ scaling first observed by Jin \textit{et al.} \cite{lishi2025}. An almost flat  W$_{\perp}$ is visible in the thermal Hall data of simple clean insulators with ballistic phonons, but not in more complex ones \cite{guo2026interactiondriventransversethermal}. This indicates that the  thermal Hall signal may have multiple sources. 

\section{Concluding remarks}
In summary, we found a thermal Hall response in quartz, which is a periodic  network of corner-sharing SiO$_4$ tetrahedra. With the same set-up, we did not find any signal in silica. The latter is composed of the same  tetrahedron-like bricks, but distorted and without long-range order. The result indicates that crystalline order is required for a thermal Hall response in solids. A comparison with a gas of neutral molecules leads us to conclude that a transverse thermal response odd in magnetic field can arise whenever the magnetic field influences entropy-producing inter-particle collisions. We found that the order of magnitude of the thermal Hall resistivity corresponds to what is expected following three assumptions: i) the heat flux generates a tiny drift; ii) a  Berry force emerges due to the coupling between this drift and the magnetic field; and iii) this force is countered by an entropic force associated with a temperature gradient. 

\section{Acknowledgments}
This work was supported by a grant attributed by the Ile de France regional council. Y.L. acknowledges a grant by China Scholarship Council.

\section{Appendix}

\renewcommand{\thefigure}{A-\arabic{figure}}
\renewcommand{\theequation}{A-\arabic{equation}}

\setcounter{figure}{0}
\setcounter{table}{0}
\setcounter{equation}{0}
\subsection{Samples and measurement technique}
Samples were obtained from SurfaceNet GmbH with dimensions of 5 $mm\times$5 $mm\times$ 0.5 $mm$. They were glued to a copper heat sink with GE Varnish (Fig. 2b). Heat flow along the sample was applied by driving a DC current through a 400 $\Omega$ resistor. The heating power was determined by multiplying the current and the voltage across the heater (monitored using a Keithley 6220). Three Cernox 1070 thermometers allowed simultaneous measurements of the longitudinal ($\Delta T_x = T_1 - T_2$) and transverse ($\Delta T_y = T_3 - T_1$) temperature differences. To minimize heat leak, 50-ohm manganin wires were used for all electrical connections from the thermometers and the heater to the puck terminals. $\Delta T_x, \Delta T_y$ and $T_1$ were derived from the resistance of the Cernox 1070 thermometers,  monitored with lock-in amplifiers (SR830) at a low frequency of f=9.777 $Hz$. To achieve higher resolution, $\Delta T_x$ and $\Delta T_y$ were measured in differential mode,  as described previously \cite{Li2020}. 

Then the longitudinal thermal conductivity $\kappa_{xx}$ and the thermal Hall conductivity $\kappa_{xy}$ can be obtained using:
\begin{equation}
    \kappa_{xx}=\frac{Q}{\Delta T_x}\frac{l}{w\cdot t}
\end{equation}
\begin{equation}
    \kappa_{xy}=\kappa_{xx}^2\frac{\Delta T_y\cdot t}{Q}
     \label{equ.2}
\end{equation}
The transverse thermal  resistivity $W_{\perp}$ is simply given by:

\begin{equation}
    W_{\perp}=\frac{\Delta T_y\cdot t}{Q}
     \label{equ.3}
\end{equation}

\subsection{Data processing }
\begin{figure*}
\centering
    \includegraphics[width=\textwidth]{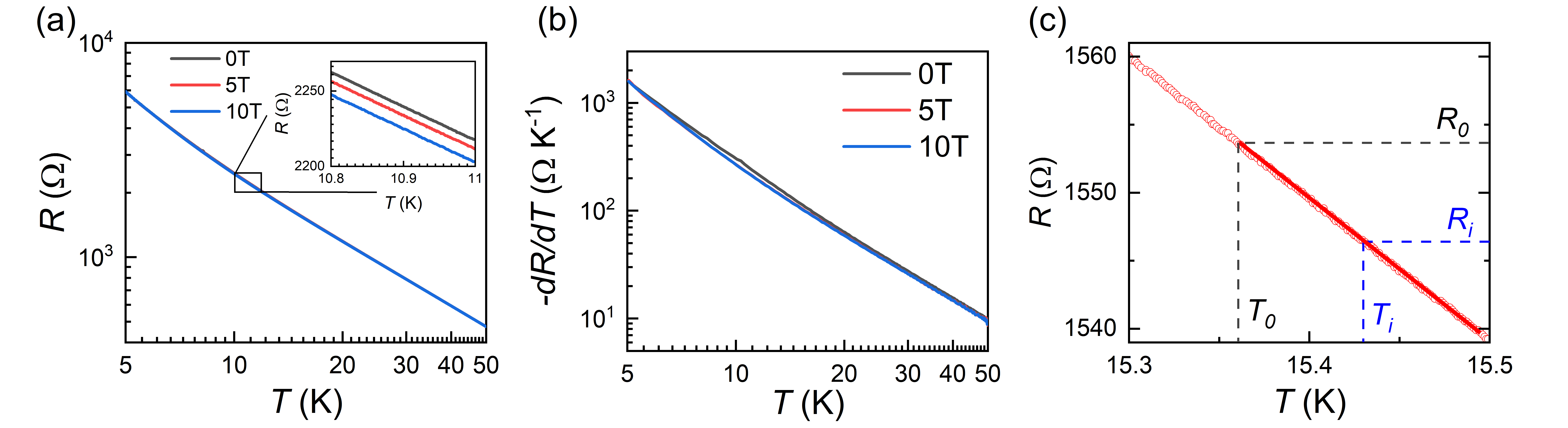}
    \caption{a) Calibration curve of Cernox 1070 in different magnetic field. The enlarged image exhibits the magneto-resistivity in Cernox 1070 thermometer. b) Temperature dependence $-dR/dT$ curve of Cernox 1070 thermometer in different magnetic field. c) Schematic diagram of fitting method for thermometer temperature.}
    \label{SM:2}
\end{figure*}

To determine the local temperature under an applied heat current, a calibration curve without heat flow is necessary. As shown in Fig. \ref{SM:2}a-b, the resistance of Cernox 1070 increases sharply with decreasing temperature enabling the detection of  $\Delta T_y$ as small as a fraction of millikelvin. The enlarged view further reveals the finite magneto-resistance of Cernox 1070, which contributes predominantly to the symmetric components in field-dependence $R_3-R_1$ curve shown in Fig. 2.

\begin{figure*}
\centering
    \includegraphics[width=0.9\textwidth]{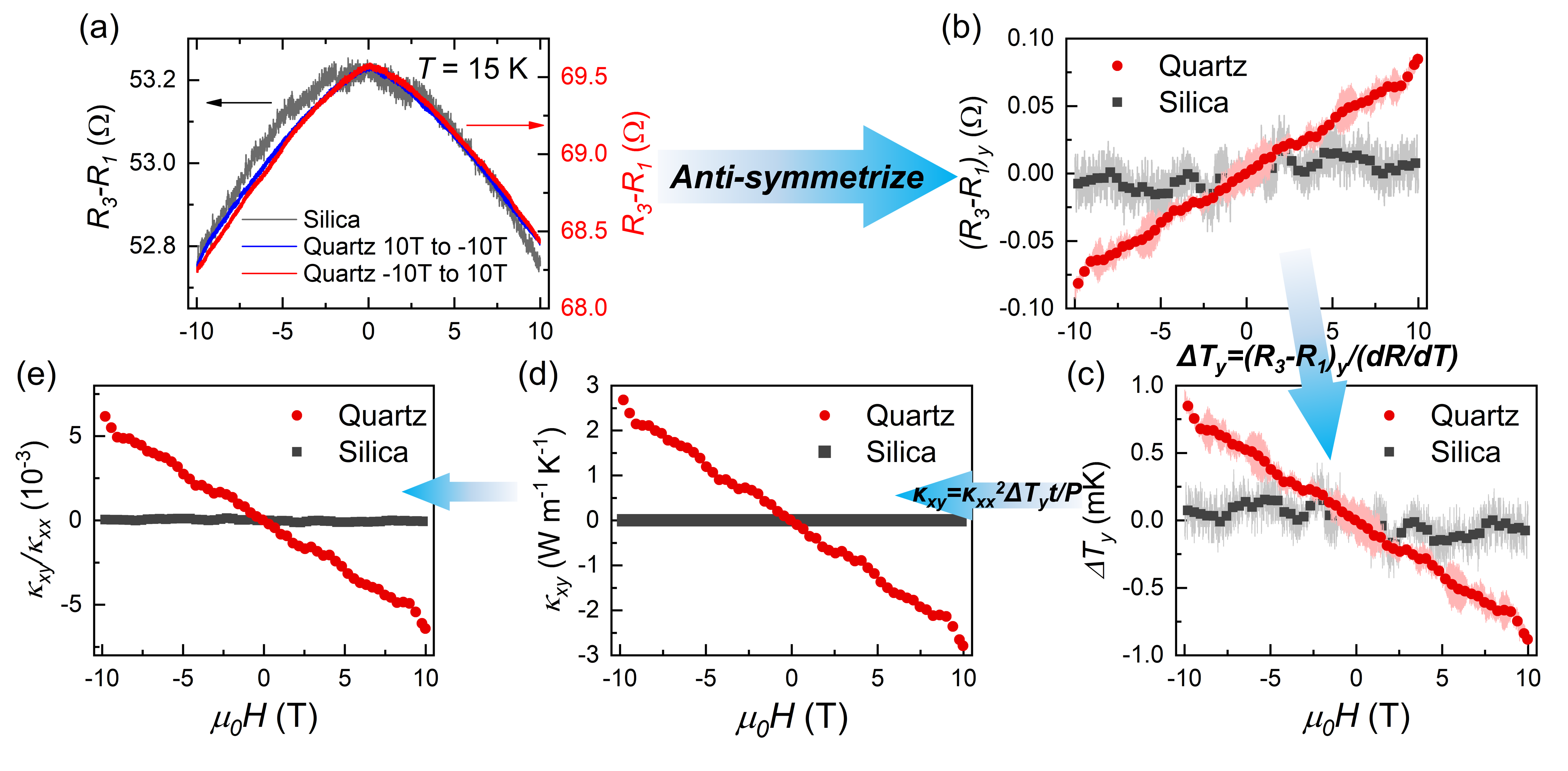}
    \caption{\textbf{Processing of the raw data.} a) Field-dependence raw data $R_3-R_1$ in quartz and silica at 15 K. b-e) Field-dependence raw data $(R_3-R_1)_y$, transverse temperature difference $\Delta T_y$, thermal Hall conductivity $\kappa_{xy}$ and thermal Hall angle $\kappa_{xy}/\kappa_{xx}$.}
    \label{SM:3}
\end{figure*}
To extract temperature from the Cernox resistance,  the measured resistance under heat flow can be fitted to the calibration curve. However, as shown in Fig. \ref{SM:2}c, the calibration curve exhibits fluctuations of the order of several 0.1 $\Omega$. This uncertainty is comparable to the resistance change corresponding to $\Delta T_y$, and thus can significantly affect the accuracy of the extracted temperature difference. To overcome this issue, we adopt a local linear approximation. Although the overall calibration curve is nonlinear, it can be regarded as linear within a narrow temperature window of approximately 0.2 K (red solid line in Fig. \ref{SM:2}c). Within this range, the local temperature can be obtained as:
\begin{equation}
    T_i=T_0-\frac{R_0-R_i}{dR/dT}
    \label{equ.4}
\end{equation}
in which $T_0$ and $R_0$ are taken from the calibration curve, $R_i$ is the measured resistance of the i-th Cernox thermometer and $dR/dT(R_0,T_0)$ is the local slope of the calibration curve.

Fig. \ref{SM:3} illustrates the raw data processing for thermal Hall conductivity $\kappa_{xy}$ in quartz and silica. In Fig. \ref{SM:3}a, the data obtained from both magnetic-field sweep directions exhibit consistent antisymmetric components, demonstrating the reproducibility of the signal and excluding spurious contributions from environmental drifts. To extract the antisymmetric component, which is attributed to the thermal Hall effect, we antisymmetrize the raw data $R_3-R_1$ as follows:
\begin{equation}
    (R_3-R_1)_y=\frac{(R_3-R_1)(+\mu_0H)-(R_3-R_1)(-\mu_0H)}{2}
    \label{equ.5}
\end{equation}

As shown in Fig. \ref{SM:3}b, a clear linear antisymmetric component is observed in quartz, whereas it is absent in silica. The transverse temperature difference can then be obtained as $\Delta T_y=(R_3-R_1)_y/(dR/dT)$. Using equation \ref{equ.2}, we  obtain the field-dependence of the thermal Hall conductivity, $\kappa_{xy}$, and  the thermal Hall angle, $\kappa_{xy}/\kappa_{xx}$, as shown in Fig. \ref{SM:3}d-e.

\bibliography{main}

@article{Zeller1971,
  title = {Thermal Conductivity and Specific Heat of Noncrystalline Solids},
  author = {Zeller, R. C. and Pohl, R. O.},
  journal = {Phys. Rev. B},
  volume = {4},
  issue = {6},
  pages = {2029--2041},
  numpages = {0},
  year = {1971},
  month = {Sep},
  publisher = {American Physical Society},
  doi = {10.1103/PhysRevB.4.2029},
  url = {https://link.aps.org/doi/10.1103/PhysRevB.4.2029}
}

@article{taraskin1997silica,
  title={Nature of vibrational excitations in vitreous silica},
  author={Taraskin, SN and Elliott, SR},
  journal={Phys. Rev. B},
  volume={56},
  number={14},
  pages={8605},
  year={1997},
  publisher={APS}
}

@article{Peierls1929,
author = {Peierls, R.},
title = {Zur kinetischen Theorie der Wärmeleitung in Kristallen},
journal = {Annalen der Physik},
volume = {395},
number = {8},
pages = {1055-1101},
  year={1929},
doi = {https://doi.org/10.1002/andp.19293950803}
}

@article{allen1999diffusons,
  title={Diffusons, locons and propagons: Character of atomic vibrations in amorphous {Si}},
  author={Allen, Philip B and Feldman, Joseph L and Fabian, Jaroslav and Wooten, Frederick},
  journal={Philosophical Magazine B},
  volume={79},
  number={11-12},
  pages={1715--1731},
  year={1999},
  publisher={Taylor \& Francis}
}

@article{Vandersande01031986,
author = {J. W. Vandersande and C. Wood},
title = {The thermal conductivity of insulators and semiconductors},
journal = {Contemporary Physics},
volume = {27},
number = {2},
pages = {117--144},
year = {1986},
publisher = {Taylor \& Francis},
doi = {10.1080/00107518608211003}

}

@article{Strohm2005,
  title = {Phenomenological Evidence for the Phonon {Hall} Effect},
  author = {Strohm, C. and Rikken, G. L. J. A. and Wyder, P.},
  journal = {Phys. Rev. Lett.},
  volume = {95},
  issue = {15},
  pages = {155901},
  numpages = {4},
  year = {2005},
  month = {Oct},
  publisher = {American Physical Society},
  doi = {10.1103/PhysRevLett.95.155901},
  url = {https://link.aps.org/doi/10.1103/PhysRevLett.95.155901}
}

@article{Sheng2006,
  title = {Theory of the Phonon {Hall} Effect in Paramagnetic Dielectrics},
  author = {Sheng, L. and Sheng, D. N. and Ting, C. S.},
  journal = {Phys. Rev. Lett.},
  volume = {96},
  issue = {15},
  pages = {155901},
  numpages = {4},
  year = {2006},
  month = {Apr},
  publisher = {American Physical Society},
  doi = {10.1103/PhysRevLett.96.155901},
  url = {https://link.aps.org/doi/10.1103/PhysRevLett.96.155901}
}

@article{Qin2012,
  title = {{Berry} curvature and the phonon {Hall} effect},
  author = {Qin, Tao and Zhou, Jianhui and Shi, Junren},
  journal = {Phys. Rev. B},
  volume = {86},
  issue = {10},
  pages = {104305},
  numpages = {9},
  year = {2012},
  month = {Sep},
  publisher = {American Physical Society},
  doi = {10.1103/PhysRevB.86.104305},
  url = {https://link.aps.org/doi/10.1103/PhysRevB.86.104305}
}

@article{Ideue2017,
  title={Giant thermal {Hall} effect in multiferroics},
  author={Ideue, T and Kurumaji, T and Ishiwata, S and Tokura, Y},
  journal={Nature Materials},
  volume={16},
  number={8},
  pages={797--802},
  year={2017},
  publisher={Nature Publishing Group},
  doi = {10.1038/nmat4905},
  url = {https://www.nature.com/articles/nmat4905#citeas}
}

@article{Sugii2017,
  title = {Thermal {Hall} Effect in a Phonon-Glass {${\mathrm{Ba}}_{3}{\mathrm{CuSb}}_{2}{\mathrm{O}}_{9}$}},
  author = {Sugii, K. and Shimozawa, M. and Watanabe, D. and Suzuki, Y. and Halim, M. and Kimata, M. and Matsumoto, Y. and Nakatsuji, S. and Yamashita, M.},
  journal = {Phys. Rev. Lett.},
  volume = {118},
  issue = {14},
  pages = {145902},
  numpages = {5},
  year = {2017},
  month = {Apr},
  publisher = {American Physical Society},
  doi = {10.1103/PhysRevLett.118.145902},
  url = {https://link.aps.org/doi/10.1103/PhysRevLett.118.145902}
}

@article{Grissonnanche2020,
  title={Chiral phonons in the pseudogap phase of cuprates},
  author={Grissonnanche, G and Th{\'e}riault, S and Gourgout, A and Boulanger, M-E and Lefran{\c{c}}ois, E and Ataei, A and Lalibert{\'e}, F and Dion, M and Zhou, J-S and Pyon, S and others},
  journal={Nature Physics},
  volume={16},
  number={11},
  pages={1108--1111},
  year={2020},
  publisher={Nature Publishing Group},
  doi = {10.1038/s41567-020-0965-y},
  url = {https://www.nature.com/articles/s41567-020-0965-y}
}

@article{Chen2020,
  title = {Enhanced Thermal {Hall} Effect in Nearly Ferroelectric Insulators},
  author = {Chen, Jing-Yuan and Kivelson, Steven A. and Sun, Xiao-Qi},
  journal = {Phys. Rev. Lett.},
  volume = {124},
  issue = {16},
  pages = {167601},
  numpages = {6},
  year = {2020},
  month = {Apr},
  publisher = {American Physical Society},
  doi = {10.1103/PhysRevLett.124.167601},
  url = {https://link.aps.org/doi/10.1103/PhysRevLett.124.167601}
}

@article{Li2020,
  title = {Phonon Thermal {Hall} Effect in Strontium Titanate},
  author = {Li, Xiaokang and Fauqu\'e, Beno\^{\i}t and Zhu, Zengwei and Behnia, Kamran},
  journal = {Phys. Rev. Lett.},
  volume = {124},
  issue = {10},
  pages = {105901},
  numpages = {6},
  year = {2020},
  month = {Mar},
  publisher = {American Physical Society},
  doi = {10.1103/PhysRevLett.124.105901},
  url = {https://link.aps.org/doi/10.1103/PhysRevLett.124.105901}
}

@article{Boulanger2020,
  title={Thermal {Hall} conductivity in the cuprate Mott insulators {Nd$_2$CuO$_4$ and Sr$_2$CuO$_2$Cl$_{2}$}},
  author={Boulanger, Marie-Eve and Grissonnanche, Ga{\"e}l and Badoux, Sven and Allaire, Andr{\'e}anne and Lefran{\c{c}}ois, {\'E}tienne and Legros, Ana{\"e}lle and Gourgout, Adrien and Dion, Maxime and Wang, CH and Chen, XH and others},
  journal={Nature Communications},
  volume={11},
  number={5325},
  pages={1--9},
  year={2020},
  publisher={Nature Publishing Group},
  doi = {10.1038/s41467-020-18881-z},
  url = {https://www.nature.com/articles/s41467-020-18881-z}
}

@article{Sim2021,
  title = {Sizable Suppression of Thermal {Hall} Effect upon Isotopic Substitution in {SrTiO$_{3}$}},
  author = {Sim, Sangwoo and Yang, Heejun and Kim, Ha-Leem and Coak, Matthew J. and Itoh, Mitsuru and Noda, Yukio and Park, Je-Geun},
  journal = {Phys. Rev. Lett.},
  volume = {126},
  issue = {1},
  pages = {015901},
  numpages = {5},
  year = {2021},
  month = {Jan},
  publisher = {American Physical Society},
  doi = {10.1103/PhysRevLett.126.015901},
  url = {https://link.aps.org/doi/10.1103/PhysRevLett.126.015901}
}

@article{Chen2022,
  title={Large phonon thermal {Hall} conductivity in the antiferromagnetic insulator {Cu$_3$TeO$_6$}},
  author={Chen, Lu and Boulanger, Marie-Eve and Wang, Zhi-Cheng and Tafti, Fazel and Taillefer, Louis},
  journal={Proceedings of the National Academy of Sciences},
  volume={119},
  number={34},
  pages={e2208016119},
  year={2022},
  publisher={National Acad Sciences},
  doi = {10.1073/pnas.2208016119},
  url = {https://www.pnas.org/doi/abs/10.1073/pnas.220801611}
}

@article{Jiang2022,
  title={Phonon drag thermal {Hall} effect in metallic strontium titanate},
  author={Jiang, Shan and Li, Xiaokang and Fauqu{\'e}, Beno{\^\i}t and Behnia, Kamran},
  journal={Proceedings of the National Academy of Sciences},
  volume={119},
  number={35},
  pages={e2201975119},
  year={2022},
  publisher={National Acad Sciences},
  doi = {10.1073/pnas.2201975119},
  url = {https://www.pnas.org/doi/abs/10.1073/pnas.2201975119}
}

@article{Uehara2022,
  title={Phonon thermal {Hall} effect in a metallic spin ice},
  author={Uehara, Taiki and Ohtsuki, Takumi and Udagawa, Masafumi and Nakatsuji, Satoru and Machida, Yo},
  journal={Nature Communications},
  volume={13},
  number={4604},
  pages={1--8},
  year={2022},
  publisher={Nature Publishing Group},
  doi = {10.1038/s41467-022-32375-0},
  url = {https://www.nature.com/articles/s41467-022-32375-0}
}

@article{Simoncelli2019,
  title={Unified theory of thermal transport in crystals and glasses},
  author={Simoncelli, Michele and Marzari, Nicola and Mauri, Francesco},
  journal={Nature Physics},
  volume={15},
  number={8},
  pages={809--813},
  year={2019},
  publisher={Nature Publishing Group},
  doi = {10.1038/s41567-019-0520-x},
  url = {https://www.nature.com/articles/s41567-019-0520-x}
}

@article{Akazawa2020,
  title = {Thermal {Hall} Effects of Spins and Phonons in Kagome Antiferromagnet {Cd-Kapellasite}},
  author = {Akazawa, Masatoshi and Shimozawa, Masaaki and Kittaka, Shunichiro and Sakakibara, Toshiro and Okuma, Ryutaro and Hiroi, Zenji and Lee, Hyun-Yong and Kawashima, Naoki and Han, Jung Hoon and Yamashita, Minoru},
  journal = {Phys. Rev. X},
  volume = {10},
  issue = {4},
  pages = {041059},
  numpages = {14},
  year = {2020},
  month = {Dec},
  publisher = {American Physical Society},
  doi = {10.1103/PhysRevX.10.041059},
  url = {https://link.aps.org/doi/10.1103/PhysRevX.10.041059}
}

@article{Flebus2022,
  title = {Charged defects and phonon {Hall} effects in ionic crystals},
  author = {Flebus, B. and MacDonald, A. H.},
  journal = {Phys. Rev. B},
  volume = {105},
  issue = {22},
  pages = {L220301},
  numpages = {6},
  year = {2022},
  month = {Jun},
  publisher = {American Physical Society},
  doi = {10.1103/PhysRevB.105.L220301},
  url = {https://link.aps.org/doi/10.1103/PhysRevB.105.L220301}
}

@article{Guo2022,
  title={Resonant thermal {Hall} effect of phonons coupled to dynamical defects},
  author={Guo, Haoyu and Joshi, Darshan G and Sachdev, Subir},
  journal={Proceedings of the National Academy of Sciences},
  volume={119},
  number={46},
  pages={e2215141119},
  year={2022},
  publisher={National Acad Sciences},
  doi = {10.1073/pnas.2215141119},
  url = {https://www.pnas.org/doi/abs/10.1073/pnas.2215141119}
}

@Article{Li2023,
author={Li, Xiaokang
and Machida, Yo
and Subedi, Alaska
and Zhu, Zengwei
and Li, Liang
and Behnia, Kamran},
title={The phonon thermal {Hall} angle in black phosphorus},
journal={Nature Communications},
year={2023},
month={Feb},
day={23},
volume={14},
number={1},
pages={1027},
issn={2041-1723},
doi={10.1038/s41467-023-36750-3},
url={https://doi.org/10.1038/s41467-023-36750-3}
}

@article{Mangeolle2022,
  title = {Phonon Thermal {Hall} Conductivity from Scattering with Collective Fluctuations},
  author = {Mangeolle, L\'eo and Balents, Leon and Savary, Lucile},
  journal = {Phys. Rev. X},
  volume = {12},
  issue = {4},
  pages = {041031},
  numpages = {13},
  year = {2022},
  month = {Dec},
  publisher = {American Physical Society},
  doi = {10.1103/PhysRevX.12.041031},
  url = {https://link.aps.org/doi/10.1103/PhysRevX.12.041031}
}

@article{sharma2024phonon,
  title = {Phonon thermal {Hall} effect in charge-compensated topological insulators},
  author = {Sharma, Rohit and Bagchi, Mahasweta and Wang, Yongjian and Ando, Yoichi and Lorenz, Thomas},
  journal = {Phys. Rev. B},
  volume = {109},
  issue = {10},
  pages = {104304},
  numpages = {9},
  year = {2024},
  month = {Mar},
  publisher = {American Physical Society},
  doi = {10.1103/PhysRevB.109.104304},
  url = {https://link.aps.org/doi/10.1103/PhysRevB.109.104304}
}

@article{Li2025,
title = {Angle-dependent planar thermal {Hall} effect by quasi-ballistic phonons in black phosphorus},
journal = {Science Bulletin},
volume = {70},
number = {12},
pages = {1962-1967},
year = {2025},
issn = {2095-9273},
doi = {https://doi.org/10.1016/j.scib.2025.03.052},
url = {https://www.sciencedirect.com/science/article/pii/S2095927325003202},
author = {Xiaokang Li and Xiaodong Guo and Zengwei Zhu and Kamran Behnia}
}

@ARTICLE{Meng2024,
       author = {{Meng}, Qingkai and {Li}, Xiaokang and {Liu}, Jie and {Zhao}, Lingxiao and {Dong}, Chao and {Zhu}, Zengwei and {Li}, Liang and {Behnia}, Kamran},
        title = "{Thermodynamic origin of the phonon Hall effect in a honeycomb antiferromagnet}",
      journal = {arXiv e-prints},
         year = 2024,
        month = mar,
          doi = {10.48550/arXiv.2403.13306},
archivePrefix = {arXiv},
       adsurl = {https://ui.adsabs.harvard.edu/abs/2024arXiv240313306M}
}

@article{Chen2024,
  title = {Planar Thermal {Hall} Effect from Phonons in Cuprates},
  author = {Chen, Lu and Le Roux, L\'ena and Grissonnanche, Ga\"el and Boulanger, Marie-Eve and Th\'eriault, Steven and Liang, Ruixing and Bonn, D. A. and Hardy, W. N. and Pyon, S. and Takayama, T. and Takagi, H. and Xu, Ke-Jun and Shen, Zhi-Xun and Taillefer, Louis},
  journal = {Phys. Rev. X},
  volume = {14},
  issue = {4},
  pages = {041011},
  numpages = {9},
  year = {2024},
  month = {Oct},
  publisher = {American Physical Society},
  doi = {10.1103/PhysRevX.14.041011},
  url = {https://link.aps.org/doi/10.1103/PhysRevX.14.041011}
}

@article{Chen2024-2,
  title={Planar thermal {Hall} effect from phonons in a {Kitaev} candidate material},
  author={Chen, Lu and Lefran{\c{c}}ois, {\'E}tienne and Vallipuram, Ashvini and Barth{\'e}lemy, Quentin and Ataei, Amirreza and Yao, Weiliang and Li, Yuan and Taillefer, Louis},
  journal={Nature Communications},
  volume={15},
  number={1},
  pages={3513},
  year={2024},
  doi = {10.1038/s41467-024-47858-5},
  publisher={Nature Publishing Group UK London}
}

@article{Ataei2024,
  title={Phonon chirality from impurity scattering in the antiferromagnetic phase of {Sr$_2$IrO$_4$}},
  author={Ataei, A and Grissonnanche, G and Boulanger, M-E and Chen, L and Lefran{\c{c}}ois, {\'E} and Brouet, V and Taillefer, L},
  journal={Nature Physics},
  volume={20},
  number={4},
  pages={585--588},
  year={2024},
  doi = {10.1038/s41567-024-02384-5},
  publisher={Nature Publishing Group UK London}
}

@Article{Behnia2025,
	title={{Phonon thermal Hall as a lattice Aharonov-Bohm effect}},
	author={Kamran Behnia},
	journal={SciPost Phys. Core},
	volume={8},
	pages={061},
	year={2025},
	publisher={SciPost},
	doi={10.21468/SciPostPhysCore.8.3.061},
	url={https://scipost.org/10.21468/SciPostPhysCore.8.3.061},
}

@article{Senftleben1930,
  author = {Senftleben, H.},
  title = {Magnetische Beeinflussung des Wärmeleitvermögens paramagnetischer Gase},
  journal = {Phys. Z},
  year = {1930},
  volume = {31},
  pages = {822--961}
}

@article{Beenakker1962,
   author = {Beenakker, J. J. M. and Scoles, G. and Knaap, H. F. P. and Jonkman, R. M.},
   title = {The influence of a magnetic field on the transport properties of diatomic molecules in the gaseous state},
   journal = {Phys. Letters},
   volume = {2},
   number = {1},
   pages = {5-6},
   ISSN = {0031-9163},
   DOI = {https://doi.org/10.1016/0031-9163(62)90091-4},
   url = {https://www.sciencedirect.com/science/article/pii/0031916362900914},
   year = {1962},
   type = {Journal Article}
}

@article{xiang2025arxiv,
  title = {Thermal {Hall} Conductivity of Semimetallic Graphite Dominated by Ambipolar Phonon Drag},
  author = {Xiang, Qiaochao and Li, Xiaokang and Guo, Xiaodong and Zhu, Zengwei and Behnia, Kamran},
  journal = {Phys. Rev. Lett.},
  volume = {136},
  issue = {5},
  pages = {056303},
  numpages = {7},
  year = {2026},
  month = {Feb},
  publisher = {American Physical Society},
  doi = {10.1103/v7z6-bztz},
  url = {https://link.aps.org/doi/10.1103/v7z6-bztz}
}

@book{Mccourt1990,
   author = {Mccourt, Frederick R W and Beenakker, Jan J M and Köhler, Walter E and Kuščer, Ivan},
   title = {Nonequilibrium Phenomena in Polyatomic Gases: Dilute Gases},
   publisher = {Oxford University Press},
   ISBN = {9780198556312},
   DOI = {10.1093/oso/9780198556312.001.0001},
   url = {https://doi.org/10.1093/oso/9780198556312.001.0001},
   year = {1990},
   type = {Book}
}

@article{zhang2010topological,
  title={Topological nature of the phonon {Hall} effect},
  author={Zhang, Lifa and Ren, Jie and Wang, Jian-Sheng and Li, Baowen},
  journal={Phys. Rev. Lett.},
  volume={105},
  number={22},
  pages={225901},
  year={2010},
  publisher={APS}
}

@article{lishi2025,
  title = {Discovery of Universal Phonon Thermal {Hall} Effect in Crystals},
  author = {Jin, X. B. and Zhang, X. and Wan, W. B. and Wang, H. R. and Jiao, Y. H. and Li, S. Y.},
  journal = {Phys. Rev. Lett.},
  volume = {135},
  issue = {19},
  pages = {196302},
  numpages = {8},
  year = {2025},
  month = {Nov},
  publisher = {American Physical Society},
  doi = {10.1103/r572-5dfm},
  url = {https://link.aps.org/doi/10.1103/r572-5dfm}
}

@article{kagan1967kinetic,
  title={Kinetic theory of gases taking into account rotational degrees of freedom in an external field},
  author={Kagan, Yu and Maksimov, L},
  journal={Sov. Phys. JETP},
  volume={24},
  pages={1272--1281},
  year={1967}
}

@article{hermans1970transverse,
  title={Transverse heat transport in polyatomic gases under the influence of a magnetic field},
  author={Hermans, LJF and Schutte, A and Knaap, HFP and Beenakker, JJM},
  journal={Physica},
  volume={46},
  number={4},
  pages={491--506},
  year={1970},
  publisher={Elsevier}
}

@article{Saito2019,
  title = {Berry Phase of Phonons and Thermal {Hall} Effect in Nonmagnetic Insulators},
  author = {Saito, Takuma and Misaki, Kou and Ishizuka, Hiroaki and Nagaosa, Naoto},
  journal = {Phys. Rev. Lett.},
  volume = {123},
  issue = {25},
  pages = {255901},
  numpages = {5},
  year = {2019},
  month = {Dec},
  publisher = {American Physical Society},
  doi = {10.1103/PhysRevLett.123.255901},
  url = {https://link.aps.org/doi/10.1103/PhysRevLett.123.255901}
}

@article{Lindsay2013,
  title = {Ab initio thermal transport in compound semiconductors},
  author = {Lindsay, L. and Broido, D. A. and Reinecke, T. L.},
  journal = {Phys. Rev. B},
  volume = {87},
  issue = {16},
  pages = {165201},
  numpages = {15},
  year = {2013},
  month = {Apr},
  publisher = {American Physical Society},
  doi = {10.1103/PhysRevB.87.165201},
  url = {https://link.aps.org/doi/10.1103/PhysRevB.87.165201}
}

@article{McCourt1967,
    author = {McCourt, F. R. and Snider, R. F.},
    title = {Thermal Conductivity of a Gas of Rotating Diamagnetic Molecules in an Applied Magnetic Field},
    journal = {The Journal of Chemical Physics},
    volume = {46},
    number = {6},
    pages = {2387-2398},
    year = {1967},
    month = {03},
    issn = {0021-9606},
    doi = {10.1063/1.1841047},
    url = {https://doi.org/10.1063/1.1841047}
}

@article{HERMANS196781,
title = {Transverse heat transport in polyatomic gases under influence of a magnetic field},
journal = {Physics Letters A},
volume = {25},
number = {2},
pages = {81-82},
year = {1967},
issn = {0375-9601},
doi = {https://doi.org/10.1016/0375-9601(67)90353-2},
url = {https://www.sciencedirect.com/science/article/pii/0375960167903532},
author = {L.J.F. Hermans and P.H. Fortuin and H.F.P. Knaap and J.J.M. Beenakker}
}

@article{Scmelcher1988,
  title = {Electronic and nuclear motion and their couplings in the presence of a magnetic field},
  author = {Schmelcher, P. and Cederbaum, L. S. and Meyer, H.-D.},
  journal = {Phys. Rev. A},
  volume = {38},
  issue = {12},
  pages = {6066--6079},
  numpages = {0},
  year = {1988},
  month = {Dec},
  publisher = {American Physical Society},
  doi = {10.1103/PhysRevA.38.6066},
  url = {https://link.aps.org/doi/10.1103/PhysRevA.38.6066}
}

@article{Yin1994,
    author = {Yin, Li and Alden Mead, C.},
    title = {Magnetic screening of nuclei by electrons as an effect of geometric vector potential},
    journal = {The Journal of Chemical Physics},
    volume = {100},
    number = {11},
    pages = {8125-8131},
    year = {1994},
    month = {06},
    issn = {0021-9606},
    doi = {10.1063/1.466806},
    url = {https://doi.org/10.1063/1.466806},
}

@article{Resta_2000,
doi = {10.1088/0953-8984/12/9/201},
url = {https://doi.org/10.1088/0953-8984/12/9/201},
year = {2000},
month = {mar},
publisher = {},
volume = {12},
number = {9},
pages = {R107},
author = {Raffaele Resta},
title = {Manifestations of {Berry's} 
phase in molecules and condensed matter},
journal = {Journal of Physics: Condensed Matter},
}

@article{Peters2022,
    author = {Peters, Laurens D. M. and Culpitt, Tanner and Tellgren, Erik I. and Helgaker, Trygve},
    title = {Magnetic-translational sum rule and approximate models of the molecular {Berry} curvature},
    journal = {The Journal of Chemical Physics},
    volume = {157},
    number = {13},
    pages = {134108},
    year = {2022},
    month = {10},
    issn = {0021-9606},
    doi = {10.1063/5.0112943},
    url = {https://doi.org/10.1063/5.0112943},

}

@article{Culpitt2021,
    author = {Culpitt, Tanner and Peters, Laurens D. M. and Tellgren, Erik I. and Helgaker, Trygve},
    title = {Ab initio molecular dynamics with screened {Lorentz} forces. {I. Calculation} and atomic charge interpretation of {Berry} curvature},
    journal = {The Journal of Chemical Physics},
    volume = {155},
    number = {2},
    pages = {024104},
    year = {2021},
    month = {07},

    issn = {0021-9606},
    doi = {10.1063/5.0055388},
    url = {https://doi.org/10.1063/5.0055388},
}

@article{xiang2025phononthermalhalleffect,
  title={Phonon thermal {Hall} effect: the roles of disorder, annealing, and metallic contacts},
  author={Xiang, Qiaochao and Li, Xiaokang and Guo, Xiaodong and Behnia, Kamran and Zhu, Zengwei},
  journal={npj Quantum Materials},
  year={2026},
  publisher={Nature Publishing Group},
  doi = {10.1038/s41535-026-00876-6},
  url = {https://doi.org/10.1038/s41535-026-00876-6}
}

@article{Jiang2025,
  title={Comment on ‘{H}igh-resolution Measurements of Thermal Conductivity Matrix and Search for Thermal {H}all Effect in  {La$_2$CuO$_4$}’},
  author={Jiang, Shan and Xiang, Qiaochao and Fauqu{\'e}, Beno{\^\i}t and Li, Xiaokang and Zhu, Zengwei and Behnia, Kamran},
  journal={arXiv:2509.14105},
  year={2025},
  doi = {https://doi.org/10.48550/arXiv.2509.14105},
}

@article{mcgaughey2019,
    author = {McGaughey, Alan J. H. and Jain, Ankit and Kim, Hyun-Young and Fu, Bo},
    title = {Phonon properties and thermal conductivity from first principles, lattice dynamics, and the {Boltzmann} transport equation},
    journal = {Journal of Applied Physics},
    volume = {125},
    number = {1},
    pages = {011101},
    year = {2019},
    month = {01},
    issn = {0021-8979},
    doi = {10.1063/1.5064602},
    url = {https://doi.org/10.1063/1.5064602}
}

@article{DeAngelis03042019,
author = {Freddy DeAngelis and Murali Gopal Muraleedharan and Jaeyun Moon and Hamid Reza Seyf and Austin J. Minnich and Alan J. H. McGaughey and Asegun Henry},
title = {Thermal Transport in Disordered Materials},
journal = {Nanoscale and Microscale Thermophysical Engineering},
volume = {23},
number = {2},
pages = {81--116},
year = {2019},
publisher = {Taylor \& Francis},
doi = {10.1080/15567265.2018.1519004},
URL = {https://doi.org/10.1080/15567265.2018.1519004}
}

@article{Hu2025,
  title={High-resolution Measurements of Thermal Conductivity Matrix and Search for Thermal {H}all Effect in {La$_2$CuO$_4$}},
  author={Hu, Jiayi and Xu, Haozhi and Yao, Juntao and Gu, Genda and Li, Qiang and Ong, NP},
  journal={arXiv:2507.21403},
  year={2025},
  doi = {https://doi.org/10.48550/arXiv.2507.21403},
}

@article{KNAAP1967643,
title = {Heat conductivity and viscosity of a gas of non-spherical molecules in a magnetic field},
journal = {Physica},
volume = {33},
number = {3},
pages = {643-670},
year = {1967},
issn = {0031-8914},
doi = {https://doi.org/10.1016/0031-8914(67)90209-1},
url = {https://www.sciencedirect.com/science/article/pii/0031891467902091},
author = {H.F.P. Knaap and J.J.M. Beenakker}
}

@article{Beenakker1970,
   author = "Beenakker, J J M and McCourt, F R",
   title = "Magnetic and Electric Effects on Transport
Properties", 
   journal= "Annual Review of Physical Chemistry",
   year = "1970",
   volume = "21",
   number = "1",
   pages = "47-72",
   doi = "https://doi.org/10.1146/annurev.pc.21.100170.000403",
   url = "https://www.annualreviews.org/content/journals/10.1146/annurev.pc.21.100170.000403",
   publisher = "Annual Reviews",
   issn = "1545-1593",
   type = "Journal Article",
  }

@article{Callaway1959,
  title = {Model for Lattice Thermal Conductivity at Low Temperatures},
  author = {Callaway, Joseph},
  journal = {Phys. Rev.},
  volume = {113},
  issue = {4},
  pages = {1046--1051},
  numpages = {0},
  year = {1959},
  month = {Feb},
  publisher = {American Physical Society},
  doi = {10.1103/PhysRev.113.1046},
  url = {https://link.aps.org/doi/10.1103/PhysRev.113.1046}
}

@article{Mizokami2018,
  title = {Lattice thermal conductivities of two {${\mathrm{SiO}}_{2}$} polymorphs by first-principles calculations and the phonon {Boltzmann} transport equation},
  author = {Mizokami, Keiyu and Togo, Atsushi and Tanaka, Isao},
  journal = {Phys. Rev. B},
  volume = {97},
  issue = {22},
  pages = {224306},
  numpages = {10},
  year = {2018},
  month = {Jun},
  publisher = {American Physical Society},
  doi = {10.1103/PhysRevB.97.224306},
  url = {https://link.aps.org/doi/10.1103/PhysRevB.97.224306}
}

@article{Kanamori1968,
author = {Kanamori, Hiroo and Fujii, Naoyuki and Mizutani, Hitoshi},
title = {Thermal diffusivity measurement of rock-forming minerals from 300° to 1100°K},
journal = {Journal of Geophysical Research (1896-1977)},
volume = {73},
number = {2},
pages = {595-605},
doi = {https://doi.org/10.1029/JB073i002p00595},
year = {1968}
}

@article{Peters2023,
author = {Peters, Laurens D. M. and Culpitt, Tanner and Tellgren, Erik I. and Helgaker, Trygve},
title = {Berry Population Analysis: Atomic Charges from the {Berry} Curvature in a Magnetic Field},
journal = {Journal of Chemical Theory and Computation},
volume = {19},
number = {4},
pages = {1231-1242},
year = {2023},
doi = {10.1021/acs.jctc.2c01138}
}

@book{Jou2009,
  author = {Jou, David and Lebon, Georgy},
  publisher = {Springer Dordrecht},
  doi = {10.1007/978-90-481-3074-0},
  title = {Extended Irreversible Thermodynamics},
  year = 2010
}

@misc{guo2026interactiondriventransversethermal,
      title={Interaction driven transverse thermal resistivity in a phonon gas}, 
      author={Xiaodong Guo and Xiaokang Li and Alaska Subedi and Zengwei Zhu and Kamran Behnia},
      year={2026},
      doi = {10.48550/arXiv.2604.03644},
      note = {arXiv preprint}, 
}

\end{document}